\documentclass[%
reprint,
superscriptaddress,
amsmath,amssymb,
aip,
]{revtex4-2}

\usepackage{graphicx}
\usepackage{dcolumn}
\usepackage{bm}
\usepackage[colorlinks,urlcolor=blue,linkcolor=blue,citecolor=blue,anchorcolor=blue]{hyperref}
\usepackage{color}

\begin{document}
	
	
	\title{Nonlinear photonic disclination states}
	
	\author{Boquan Ren}
	\affiliation{Key Laboratory for Physical Electronics and Devices of the Ministry of Education \& Shaanxi Key Lab of Information Photonic Technique, School of Electronic Science and Engineering, Faculty of Electronic and Information Engineering, Xi'an Jiaotong University, Xi'an 710049, China}
	
	\author{Hongguang Wang}
	\affiliation{Key Laboratory for Physical Electronics and Devices of the Ministry of Education \& Shaanxi Key Lab of Information Photonic Technique, School of Electronic Science and Engineering, Faculty of Electronic and Information Engineering, Xi'an Jiaotong University, Xi'an 710049, China}
	
	\author{Yaroslav V. Kartashov}
	\affiliation{Institute of Spectroscopy, Russian Academy of Sciences, Troitsk, Moscow, 108840, Russia}
	
	\author{Yongdong Li}
	\affiliation{Key Laboratory for Physical Electronics and Devices of the Ministry of Education \& Shaanxi Key Lab of Information Photonic Technique, School of Electronic Science and Engineering, Faculty of Electronic and Information Engineering, Xi'an Jiaotong University, Xi'an 710049, China}
	
	\author{Yiqi Zhang}
	\email{zhangyiqi@xjtu.edu.cn}
	\affiliation{Key Laboratory for Physical Electronics and Devices of the Ministry of Education \& Shaanxi Key Lab of Information Photonic Technique, School of Electronic Science and Engineering, Faculty of Electronic and Information Engineering, Xi'an Jiaotong University, Xi'an 710049, China}
	
	\date{\today}
	
	\begin{abstract}
		\noindent
		Higher-order topological insulators are unusual materials that can support topologically protected states, whose dimensionality is lower than the dimensionality of the structure at least by 2. Among the most intriguing examples of such states are zero-dimensional corner modes existing in two-dimensional higher-order insulators. In contrast to corner states, recently discovered disclination states also belong to the class of higher-order topological states, but are bound to the boundary of the disclination defect of the higher-order topological insulator and can be predicted using the bulk-disclination correspondence principle. Here, we present the first example of the nonlinear photonic disclination state bifurcating from its linear counterpart in the disclination lattice with a pentagonal or heptagonal core. We show that nonlinearity allows to tune location of the disclination states in the bandgap and notably affects their shapes. The structure of the disclination lattice is crucial for stability of these nonlinear topological states: for example, disclination states are stable in the heptagonal lattice and are unstable nearly in the entire gap of the pentagonal lattice. Nonlinear disclination states reported here are thresholdless and can be excited even at low powers. Nonlinear zero-energy states coexisting in these structures with disclination states are also studied. Our results suggest that disclination lattices can be used in the design of various nonlinear topological functional devices, while disclination states supported by them may play an important role in applications, where strong field confinement together with topological protection are important, such as the design of topological lasers and enhancement of generation of high harmonics.
		
	\end{abstract}
	
	\maketitle
	

	\section{Introduction}
	
	Topological insulators that behave as conventional insulators in the bulk, but allow topologically protected currents at their edges are frequently considered as a new phase of matter~\cite{hasan.rmp.82.3045.2010,qi.rmp.83.1057.2011}. Originating from solid state physics, the concept of topological insulators nowadays extends to many other areas, such as photonics~\cite{haldane.prl.100.013904.2008, wang.nature.461.772.2009, rechtsman.nature.496.196.2013, lindner.np.7.490.2011, hafezi.np.7.907.2011, stuetzer.nature.560.461.2018, yang.nature.565.622.2019, maczewsky.science.370.701.2020, biesenthal.science376.eabm2842.2022, pyrialakos.nm.21.634.2022, barik.science.359.666.2019, mittal.nature.561.502.2018}, acoustics~\cite{yang.prl.114.114301.2015, he.np.12.1124.2016, lu.np.13.369.2017, ma.nrp.1.281.2019}, physics of atomic systems~\cite{jotzu.nature.515.237.2014}, 
	mechanics~\cite{huber.np.12.621.2016}, and physics of polariton condensates in microcavities~\cite{nalitov.prl.114.116401.2015, klembt.nature.562.552.2018, liu.science.370.600.2020}, to name just a few. 
	After experimental realization of the first photonic topological insulator~\cite{wang.nature.461.772.2009} and first Floquet photonic insulator~\cite{rechtsman.nature.496.196.2013}, the interest to rich physics of photonic topological systems has grown dramatically and nowadays one can talk about the birth of a new area -- topological photonics~\cite{lu.np.8.821.2014, ozawa.rmp.91.015006.2019}.
Among the reasons of the blooming of this research direction are not only unusual physical properties of photonic topological systems, but also richness of their potential practical applications, such as the development of robust topological lasers~\cite{jean.np.11.651.2017, bandres.science.359.eaar4005.2018, bahari.science.358.636.2017, parto.prl.120.113901.2018, zhao.nc.9.981.2018, kartashov.prl.122.083902.2019, zeng.nature.578.246.2020},
telecommunications and robust information transfer~\cite{shalaev.nn.14.31.2019,yang.np.14.446.2020},
and design of on-chip topological devices~\cite{chen.prl.126.230503.2021, yang.ap.4.046002.2022}.

Formation of topological edge states is usually connected with breakup of certain symmetries of the Hamiltonian of the system. For example, systems with broken time-reversal symmetry can support unidirectional edge states, while in valley-Hall systems topological edge states appear, when inversion symmetry of the structure is broken~\cite{lu.np.8.821.2014, ozawa.rmp.91.015006.2019}. Particular attention is now paid to so-called higher-order topological insulator (HOTI) phases that usually can be realized by proper shift of the sites in the unit cell of the structure~\cite{peterson.nature.555.346.2018, xue.nm.18.108.2019, ni.nm.18.113.2019, chen.prl.122.233902.2019, xie.prl.122.233903.2019, mittal.np.13.692.2019, serra.nature.555.342.2018, noh.np.12.408.2018, wang.pr.9.1854.2021, imhof.np.14.925.2018, zhang.np.15.582.2019, zangeneh.prl.123.053902.2019, qi.prl.124.206601.2020, hassan.np.13.697.2019, xu.sb.66.1740.2021,li.sb.67.2040.2022,li.sb.67.2040.2022,zheng.sb.67.2069.2022}, 
and whose distinctive feature is the existence of topologically protected states with dimensionality at least by 2 lower than the dimensionality of the underlying structure. For example, two-dimensional (2D) HOTI along with 1D topological edge states supports also 0D corner modes~\cite{xie.nrp.3.520.2021}. By analogy with bulk-boundary correspondence principle explaining the emergence of the edge states in conventional topological insulators~\cite{ozawa.rmp.91.015006.2019}, higher-order bulk-boundary correspondence was proposed to explain the emergence of the corner states in HOTIs, which suggests that the corner state is a result of the filling anomaly associated with the fractional charge~\cite{peterson.science.368.1114.2020}. Corner states in HOTIs are characterized by strong concentration of the light field, beneficial for the enhancement of nonlinear processes~\cite{kruk.nl.21.4592.2021} and realization of low-threshold lasing~\cite{zhang.light.9.109.2020, kim.nc.11.5758.2020, han.acs.7.2027.2020, zhong.apl.6.040802.2021}, as well as for construction of high-Q topological cavities~\cite{ota.optica.6.786.2019, xie.lpr.14.1900425.2020}.

It was proposed that disclinations in topological crystalline insulators may lead to the emergence of very unusual type of topological states~\cite{ruegg.prl.110.046401.2013, teo.prl.111.047006.2013, banlcazar.prb.89.224503.2014, benalcazar.prb.99.245151.2019, li.prb.101.115115.2020, wu.pr.9.668.2021}.
The disclination that serves as a crystallographic defect, disrupts the lattice structure and strongly traps fractional charges, alongside which higher-order topological states can be predicted. The bulk-disclination correspondence proposed for such systems illustrates the importance of fractional charges as a probe of crystalline topology. Very recently, fractional charges as well as the formation of the higher-order topological states due to the disclination defects were observed experimentally in linear systems~\cite{peterson.nature.589.376.2021, liu.nature.589.381.2021, chen.prl.129.154301.2022}. These higher-order topological states are \textit{disclination states} and they typically localize at the boundary of the hollow disclination core of the structure. In addition, disclination defects are useful for vortex state generation~\cite{wang.nc.12.3654.2021} and may act as a domain wall hosting photonic topological edge states~\cite{wang.prl.124.243602.2020}. If the disclination defect does not break the chiral symmetry of the system, midgap zero-energy states can also form in the system, which are confined to the disclination core~\cite{deng.prl.128.174301.2022} even though there is no fractional charge.

It is well known that one of the advantages of the photonic platform is the possibility of strong nonlinear response of the system.
However, the impact of nonlinearity on topological systems with disclinations and on topological disclination states, in particular, have never been explored. At the same time, the nonlinearity plays a significant role in control and switching of topological states~\cite{smirnova.apr.7.021306.2020}, and brings a variety of intriguing phenomena, including modulational instability~\cite{kartashov.optica.3.1228.2016} and bistability of the unidirectional edge states \cite{kartashov.prl.119.253904.2017}, topological transitions~\cite{hadad.ne.1.178.2018, maczewsky.science.370.701.2020}, formation of topological solitons~\cite{lumer.prl.111.243905.2013, mukherjee.science.368.856.2020, leykam.prl.117.143901.2016, ablowitz.pra.96.043868.2017, ivanov.acs.7.735.2020, zhong.ap.3.056001.2021}, and breakdown of topological transport~\cite{marius.nature.596.63.2021, fu.prl.128.154101.2022}. It is thus important to understand how nonlinear response of the system can change the properties, localization, and conditions of formation of topological disclination states.

\begin{figure*}[htbp]
\centering
\includegraphics[width=\textwidth]{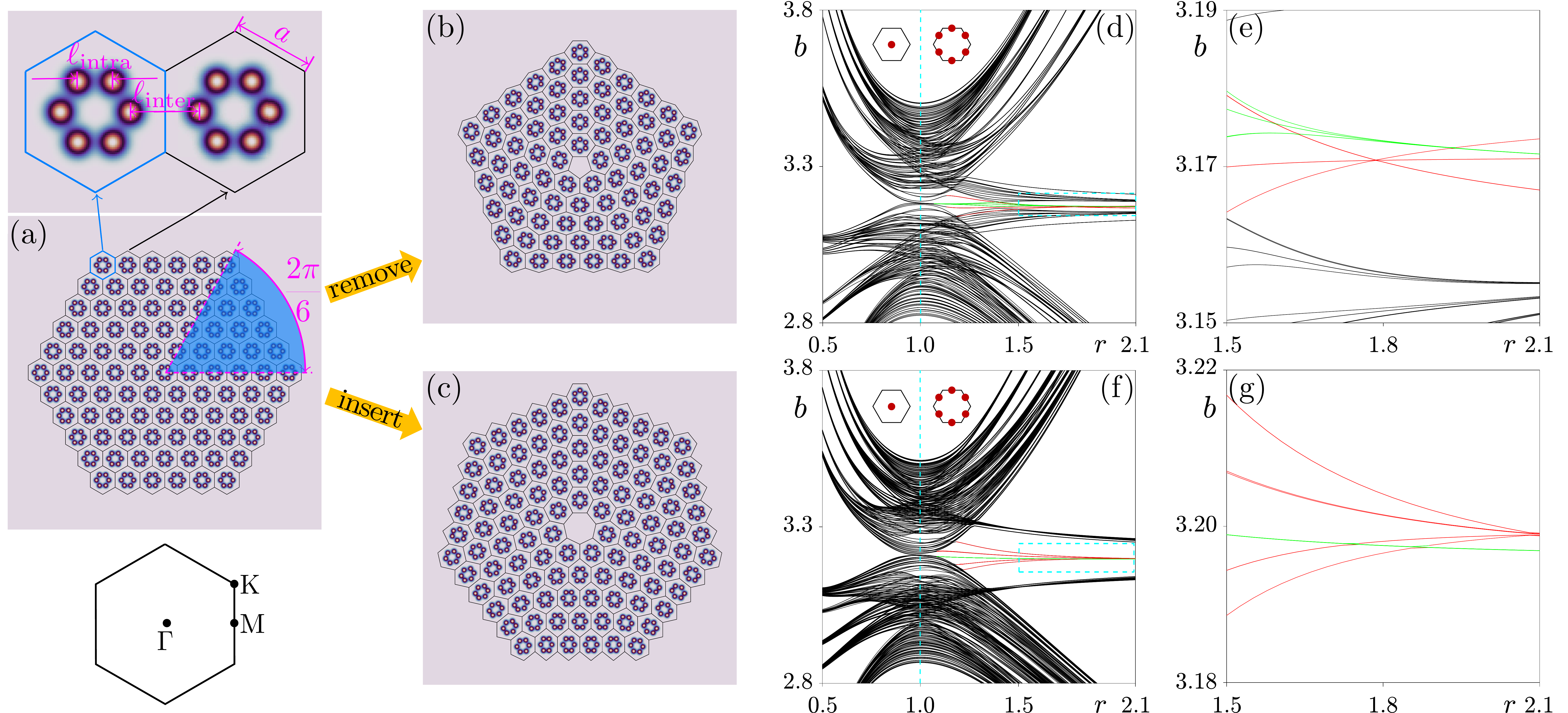}
\caption{\textbf{Disclination lattices and corresponding linear spectra.}
	(a) Illustration of the principle of construction of the disclination lattices with pentagonal (b) and heptagonal (c) cores from the honeycomb lattice by removing or inserting a sector with the Frank angle $\Omega=2\pi/6$, respectively. Two magnified unit cells displayed above panel (a) indicate spacing $\ell_{\rm intra}$ and $\ell_{\rm inter}$ between the waveguides when shift is introduced, while the first Brillouin zone of the original honeycomb structure with high-symmetric points $(\Gamma,\,\rm K,\, M)$ is shown below panel (a). The quantity $r=\ell_{\rm intra}/\ell_{\rm inter}$ characterizes the separation between sites in the unit cell. (d) Spectrum of the disclination lattice with pentagonal core versus $r$ in which the critical value $r_c=1$ is indicated by the vertical dashed line. Wannier centers (red dots) are also shown, which are at the center of the unit cell if $r<r_c$ and at the edges of the unit cell if $r>r_c$. (e) Magnified band structure with red and green lines indicating disclination and zero-energy states, respectively. (f,\,g) Spectrum and its magnified version for the lattice with a heptagonal core.}
\label{fig1}
\end{figure*}

\begin{figure*}[htpb]
\centering
\includegraphics[width=\textwidth]{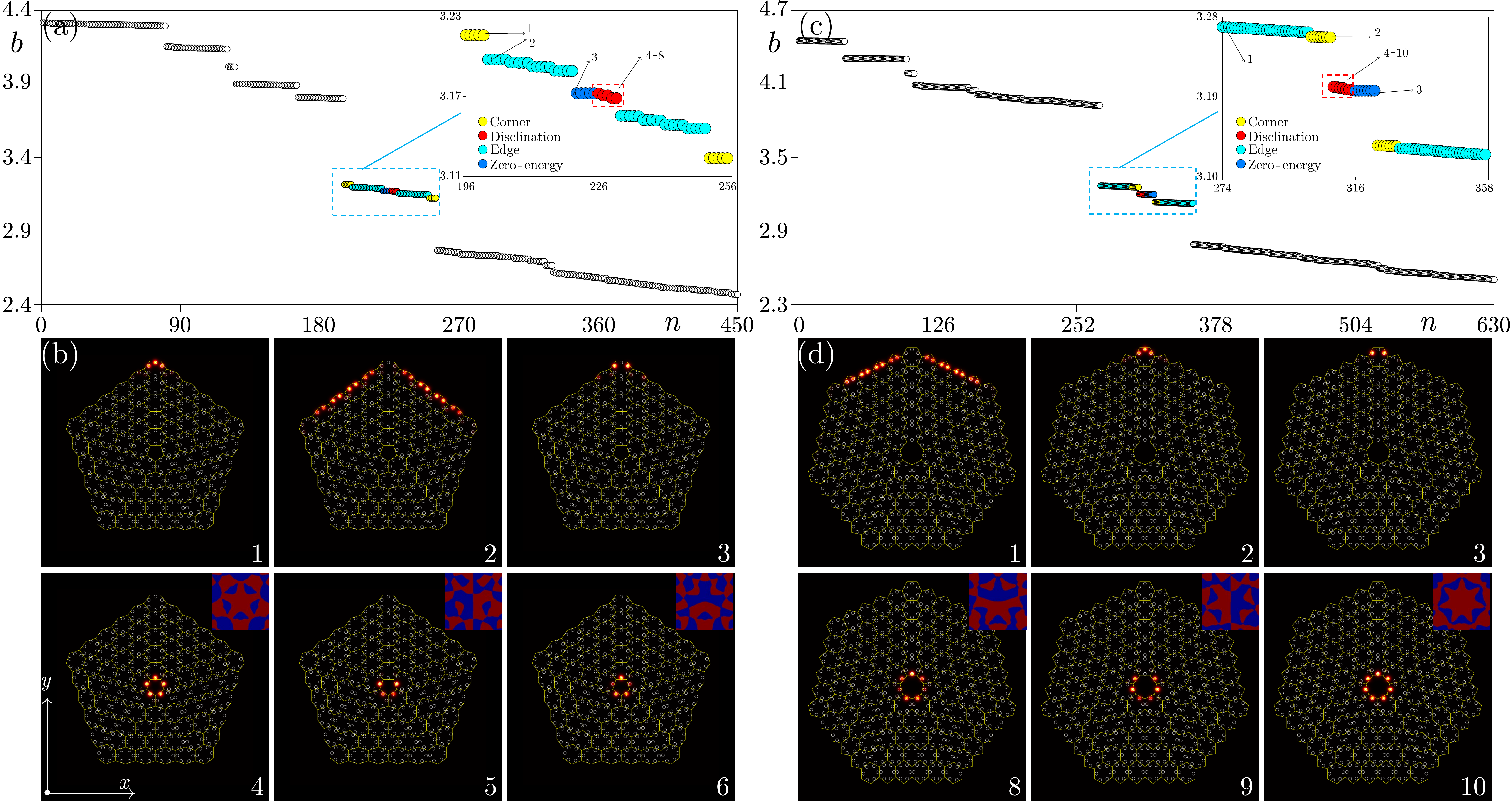}
\caption{\textbf{Examples of the disclination states.}
	(a) Eigenvalues of the modes of the disclination lattice with a pentagonal core at $r=1.92$. Yellow, cyan, blue, and red dots correspond to the corner, edge, zero-energy, and disclination states, respectively. Black dots are bulk states. (b) Field modulus distributions of exemplary states shown within the window $-35\le x,y\le 35$. Deformed lattice cells are shown with yellow lines, while lattice sites are indicated by white circles. (c,\,d) Eigenvalues of modes and examples of their profiles for the lattice with a heptagonal core at $r=1.92$. The insets in the bottom panels show phase distributions in the disclination states (red color represents $\pi$ and blue color $0$) within the window $-10\le x,y \le 10$.}
\label{fig2}
\end{figure*}

In this work, for the first time to our knowledge, we report on the formation and stability of the nonlinear disclination states in photonic structures with different symmetries. Based on the honeycomb lattice, we construct two different photonic lattices with a pentagonal disclination core and a heptagonal disclination core by removing or inserting a sector with a Frank angle of $2\pi/6$, respectively. The chiral symmetry of both these disclination lattices is broken and fractional charge will appear in the unit cell around the disclination core if they are in the topological insulator phase. We present spectra of modes of such disclination lattices and obtain disclination states in the topological phase. We found that in the presence of the focusing nonlinearity of the material the nonlinear thresholdless disclination states can bifurcate from linear ones and that their propagation constants can shift under the action of nonlinearity into the gap or even further into allowed band leading to notable modification of the shape of the state. Linear stability analysis predicts important differences in stability properties of the nonlinear disclination states in lattices with pentagonal and heptagonal cores. Importantly, all our results are obtained in the frames of the continuous model (nonlinear Schr\"odinger-like equation describing paraxial light propagation) taking into account all details of shallow refractive index landscape and couplings between all waveguides forming the lattice.
The nonlinear disclination state opens up many potentials for deep exploration of the
behavior and applications of higher-order topological states.

\begin{figure*}[htpb]
\centering
\includegraphics[width=\textwidth]{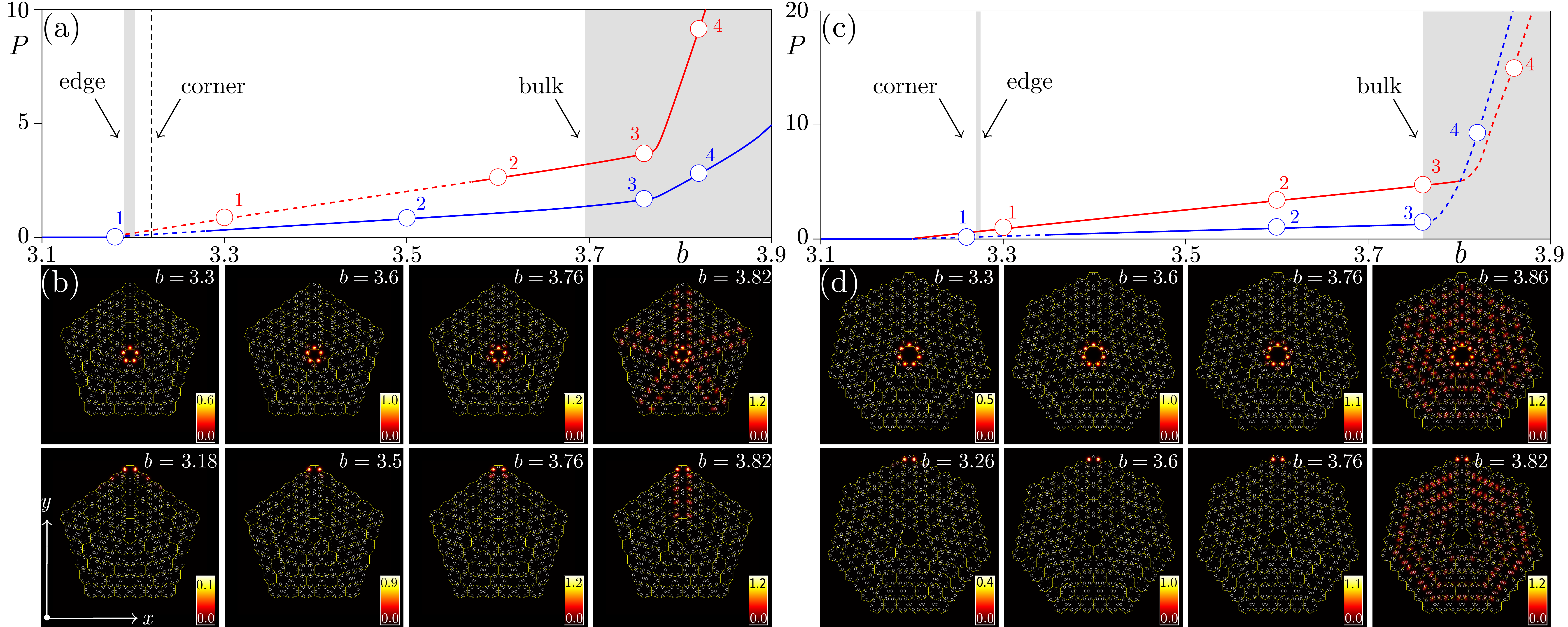}
\caption{\textbf{Nonlinear disclination and zero-energy state families.}
	(a) Nonlinear families of topological states in the disclination lattice with a pentagonal core at $r=1.82$. Red line corresponds to the disclination state, while blue line corresponds to the zero-energy one. Solid lines correspond to stable branches, while dashed lines correspond to unstable ones. Gray regions represent the bulk band (right) and edge band (left), while vertical dashed line shows band of corner states. (b) Field modulus distributions at different propagation constant values corresponding to the dots numbered $1\sim4$ in panel (a). Panels in the top row show disclination states and those in the bottom row show zero-energy states. (c,\,d) Setup as in (a,\,b), but for the disclination lattice with a heptagonal core at $r=1.79$.}
\label{fig3}
\end{figure*}

\section{Disclination lattice, fractional disclination charge and disclination state}

We consider paraxial propagation of a light beam in the material with imprinted waveguide array and focusing cubic nonlinearity that can be described by the nonlinear Schr\"odinger-like equation for the dimensionless light field amplitude $\psi$:
\begin{equation}\label{eq1}
i \frac{\partial \psi}{\partial z}=-\frac{1}{2} \left( \frac{\partial^2}{\partial x^2} + \frac{\partial^2}{\partial y^2} \right) \psi
-\mathcal{R}(x,y) \psi-|\psi|^{2} \psi.
\end{equation}
Here, $x,y$ are the transverse coordinates normalized to the characteristic transverse scale $r_0=10\,\mu \rm m$, $z$ is the propagation distance normalized to the diffraction length $kr_0^2$, where ${k=2\pi n/\lambda}$ is the wavenumber, ${\lambda=800\,\rm nm}$ is the wavelength, $n=1.45$ is the refractive index of the material, the function ${\mathcal R}(x,y)$ describes the waveguide array and it can be written as
\begin{equation}
{\mathcal R}(x,y) = p \sum_{m,n} e^{-[(x-x_m)^2 + (y-y_n)^2]/\sigma^2},
\end{equation}
where $(x_m,y_n)$ are the coordinates of the lattice ``sites'' (i.e. waveguides), $\sigma$ is the waveguide width, and $p=k^2r_0^2\delta n/n$ is the depth of the lattice that is proportional to the refractive index modulation depth $\delta n$. We assume here that the structure that we consider is fabricated using the fs-laser direct writing technique in fused silica~\cite{kirsch.np.17.995.2021, kartashov.prl.128.093901.2022}.
The original finite honeycomb lattice that is used for the construction of disclination structures is displayed in Fig.~\ref{fig1}(a). The horizontal separation between centers of two neighboring honeycomb cells (indicated by solid hexagons) in this non-deformed structure is $\sqrt{3}a$, where $a$ determines the size of the unit cell as marked in Fig.~\ref{fig1}(a). We further use the parameters $a=3.2$ (corresponding to $32\,\mu\rm m$), $\sigma=0.5$ (corresponding to $5\, \mu \rm m$) and ${p=10}$ (${\delta n \sim 1.12\times 10^{-3}}$) representative for fs-laser written waveguide arrays~\cite{tan.ap.3.024002.2021, li.ap.4.024002.2022}. In our normalizations dimensionless propagation distance $z=1$ corresponds to $\sim 1.14~\textrm{mm}$, while dimensionless intensity $|\psi|^2=1$ corresponds to peak intensity of $I=n|\psi|^2/k^2r_0^2n_2 \approx 5 \times 10^{15} ~\textrm{W/m}^2$. Similar structures can be also optically induced in photorefractive crystals~\cite{fu.np.14.663.2020, hu.light.10.164.2021, zhong.ap.3.056001.2021}, where the nonlinearity is saturable. Notice that continuous nonlinear Schrödinger equation (\ref{eq1}) takes into account all details of the refractive index landscape created in the material and on this reason it accounts for coupling between all waveguides in the structure as well as for dynamics of modal fields inside individual waveguides.

In each unit cell of the lattice there are six sites which form a hexagon with the center in the center of the unit cell. The spacing between sites in the unit cell can be increased or reduced. This gives rise to the structure with separation between two neighboring sites in the unit cell denoted as $\ell_{\rm intra}$ and horizontal separation between two nearest sites from different cells denoted as $\ell_{\rm inter}$~\cite{wu.prl.114.223901.2015,yves.nc.8.16023.2017}, as illustrated in Fig.~\ref{fig1}(a). The case $\ell_{\rm intra} = \ell_{\rm inter}=a$ corresponds to the ideal non-deformed honeycomb structure. 
Further, by removing a sector with the Frank angle $\Omega=2\pi/6$ that is shaded in Fig.~\ref{fig1}(a), and gluing the remaining nodes of the model, i.e. filling the remaining gap by uniformly stretching the lattice in the circular direction, one obtains a finite disclination lattice with a pentagonal core [see Fig.~\ref{fig1}(b)]. Similarly, one may also insert an extra sector to the honeycomb lattice and adjust it uniformly in the circular direction (by compressing the lattice) to obtain a disclination lattice with a heptagonal core, as shown in Fig.~\ref{fig1}(c).

To explore the spectrum of such finite disclination lattices, we neglect the nonlinear term in Eq.~(\ref{eq1}) and search for the eigenmodes of the structure in the form $\psi=u(x,y) e^{ibz}$, with $b$ being the propagation constant (eigenvalue) and $u(x,y)$ being the modal field. In doing so, we obtain from Eq.~(\ref{eq1}) the linear eigenvalue problem 
\begin{equation}
bu=\frac{1}{2} \left( \frac{\partial^2}{\partial x^2} + \frac{\partial^2}{\partial y^2} \right)u + \mathcal{R}u,
\end{equation} 
that can be solved numerically by using the plane-wave expansion method. In Fig.~\ref{fig1}(d) we present the spectrum of the modes of disclination lattice with the pentagonal core as a function of parameter $r=\ell_{\rm intra}/\ell_{\rm inter}$ determining the intra- and inter-cell separation between waveguides. One finds that topological states appear in the forbidden gap of the spectrum in the region $r>r_c=1$, with $r=r_c$ indicated in the figure by the vertical dashed line. In Fig.~\ref{fig1}(e) we plot the magnified part of the spectrum that is indicated by the dashed rectangle in Fig.~\ref{fig1}(d). One can observe the formation of the disclination states shown by the red lines, and of the ``zero-energy'' states (we keep this traditional notion for them that came from tight-binding models also here despite the fact that their propagation constant can actually change with $r$ in our continuous system) shown by the green lines. One can see that disclination states in general are not degenerate, except for the point $r\approx1.79$. The spectrum of the disclination lattice with a heptagonal core is presented in Fig.~\ref{fig1}(f), while its magnified version is shown in Fig.~\ref{fig1}(g). The formation of disclination states in this lattice is obvious too. Notice that localization of disclination states generally increases with the increase of parameter $r$. The emergence of disclination states is associated with transition of the lattice into topologically nontrivial phase when parameter $r$ exceeds $r_c$, as described by the associated topological invariant (see \textbf{Methods}). 

Disclination lattices support several types of modes. To illustrate them we choose a particular value of parameter $r=1.92$ exceeding the value $r_c=1$ (i.e. the lattice is in topological insulator phase) and show in Fig.~\ref{fig2}(a) for the lattice with a pentagonal core the eigenvalues of $n=450$ modes with largest propagation constants. In this figure corner, edge, zero-energy, and disclination states are highlighted by the colour dots, while bulk states are indicated by the black dots. Field modulus distributions of some modes are depicted in Fig.~\ref{fig2}(b), where they are superimposed on the lattice profile with indicated lattice cells (yellow lines) and sites (white circles). One finds that both corner and zero-energy states form around the outer corners of the finite lattice, i.e. they belong to the class of 0D states. Notice that in contrast to corner states, zero-energy ones do not occupy the very corner site of the lattice. In our sufficiently large lattice there are forty edge states (cyan dots), ten corner states (yellow dots), and five degenerated zero-energy states (blue dots) in the lattice with a pentagonal core. Beside these states that can also be encountered in usual HOTIs~\cite{noh.np.12.408.2018, hassan.np.13.697.2019, ni.nm.18.113.2019, xue.nm.18.108.2019, chen.prl.122.233902.2019, xie.prl.122.233903.2019},
there are also five disclination states associated with the fractional disclination charge (see \textbf{Methods}), which are labeled by the red dots with numbers $4\sim 8$. We display the first three disclination states (numbered $4\sim6$) in the second row of Fig.~\ref{fig2}(b). In contrast to all other states, disclination states localize at the five sites of the disclination core. As mentioned above, five disclination states are not completely degenerate: only pair of states with numbers 5 and 6 and pair of states 7 and 8 are degenerate ones. Degeneracy for such states is lifted because they are localized on compact inner disclination core, where sites located on the core do feel neighboring sites belonging to the core and therefore interact. We also depict typical phase distributions for disclination states in the insets of the second row of Fig.~\ref{fig2}(b). The disclination state numbered 4 with the largest propagation constant includes five in-phase spots, while phase distributions in all other disclination states are more complex, with alternating groups of in-phase and out-of-phase spots.

\begin{figure}[htpb]
	\centering
	\includegraphics[width=\columnwidth]{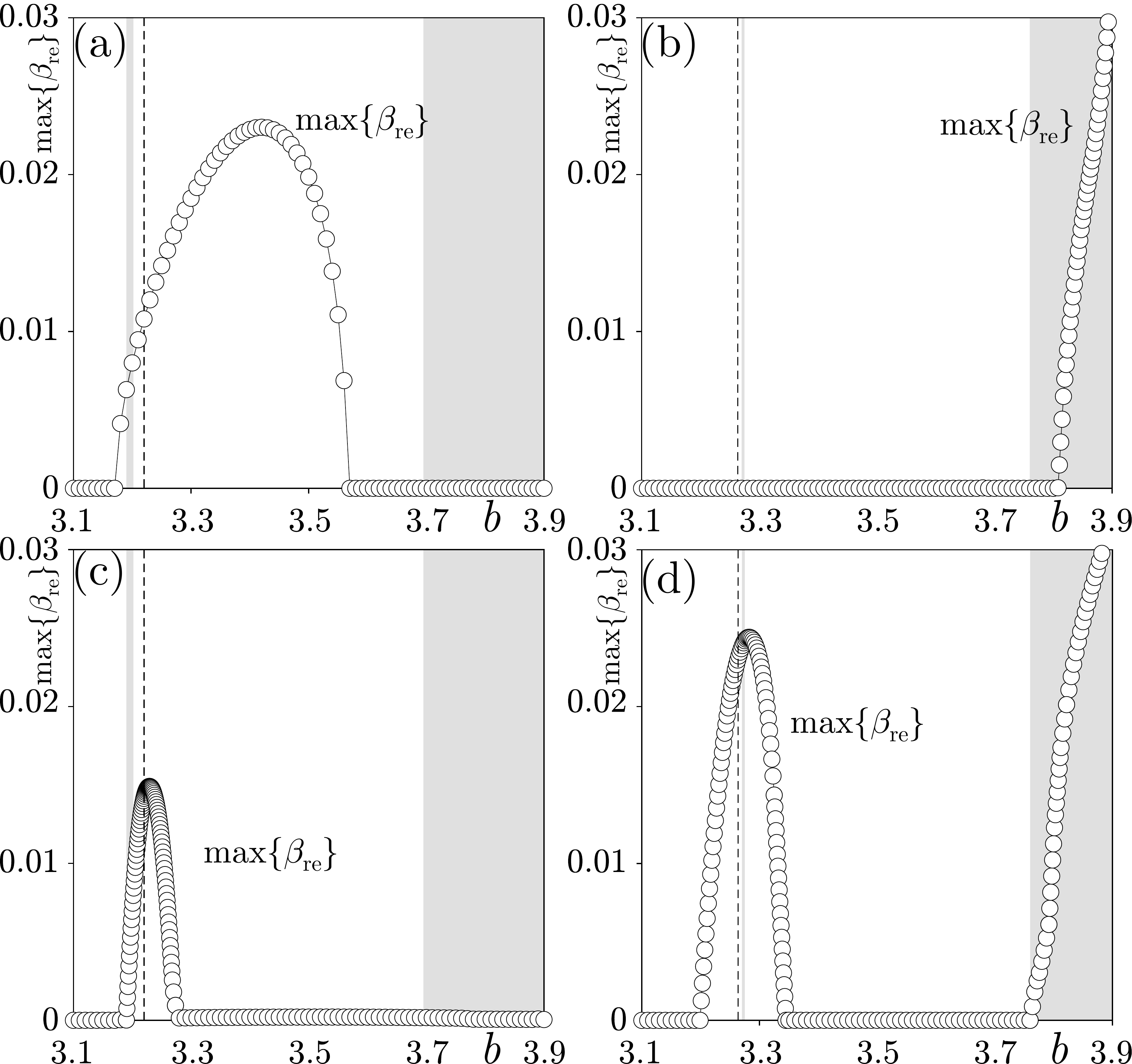}
	\caption{\textbf{Stability properties of the disclination and zero-energy states.}
		Maximal real part of the perturbation growth rate $\beta$ versus propagation constant $b$ for the disclination state (a,\,b) and zero-energy state families (c,\,d) in the lattice with a pentagonal core (a,\,c) and a heptagonal core (b,\,d). Parameters are the same as in Fig.~\ref{fig3}.}
	\label{fig4}
\end{figure}

Disclination lattice with a heptagonal core also supports corner, edge, zero-energy, and disclination states, as illustrated in Fig.~\ref{fig2}(c) showing eigenvalues of modes at $r=1.92$. Field modulus distributions in representative states are illustrated in Fig.~\ref{fig2}(d). According to the symmetry of this disclination lattice, there are fifty-six edge states (cyan dots), fourteen corner states (yellow dots), seven degenerate zero-energy states (blue dots), and seven disclination states (red dots). Again, the disclination states are also not completely degenerate. 
In this particular case, pair states numbered 4 and 5, pair states numbered 6 and 7, and pair states numbered 8 and 9 depicted in Fig.~\ref{fig2}(d) are degenerate. Notice that the state numbered 10 consists of seven in-phase spots. In the following section, we will investigate nonlinear disclination states bifurcating from the linear disclination state numbered 4 for the pentagonal case and state numbered 10 for the heptagonal case.

\begin{figure*}[htpb]
	\centering
	\includegraphics[width=\textwidth]{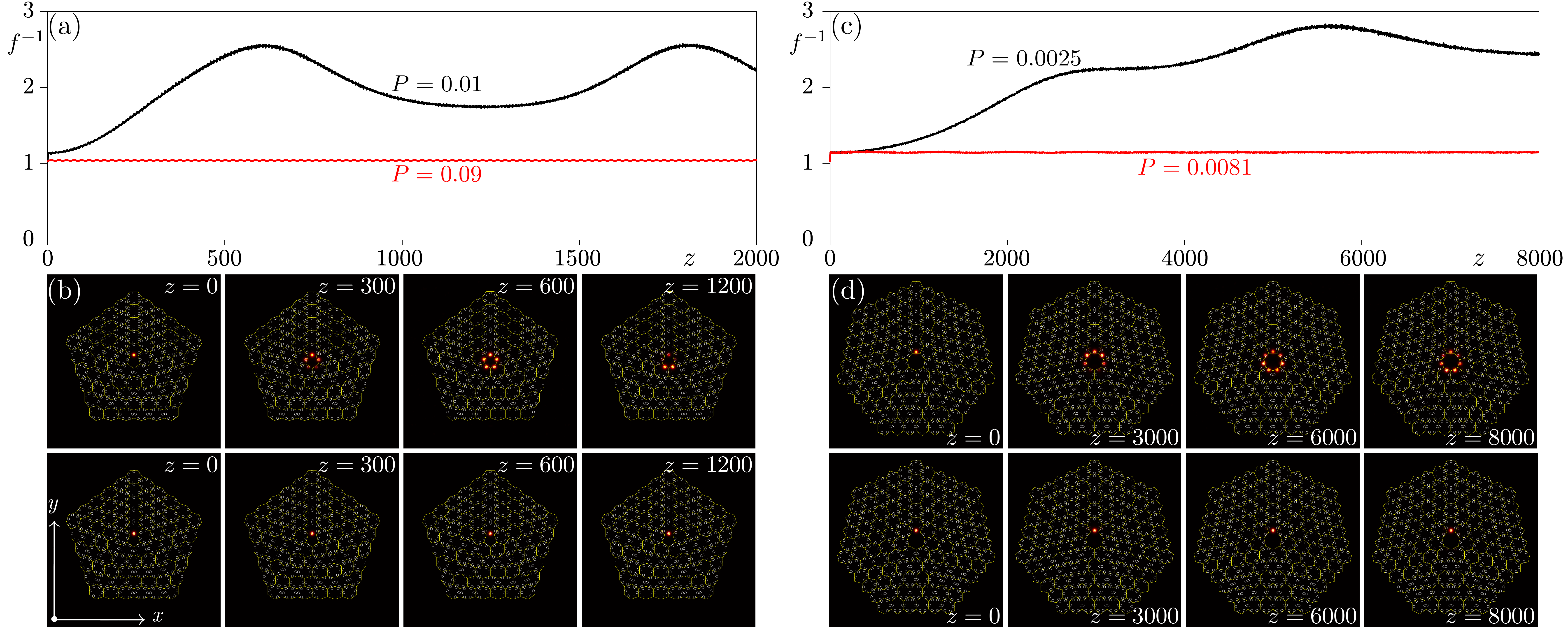}
	\caption{\textbf{Dynamical excitation of the nonlinear disclination states.}
		(a) The inverse form-factor $f^{-1}$ of the field distribution versus propagation distance for input Gaussian beams with different powers injected into single site of the pentagonal disclination core at $r=2.17$. (b) Field modulus distributions at different distances. Top and bottom panels correspond to the black and red curves in (a), respectively. (c,\,d) Setup as in (a,\,b), but for the structure with heptagonal core.}
	\label{fig5}
\end{figure*}

\section{Nonlinear disclination states}

We now take into account focusing nonlinearity of the underlying material and consider nonlinear disclination and zero-energy state families $\psi=u(x,y) e^{ibz}$ bifurcating from corresponding linear states. 
Note that similar phenomena can be expected if defocusing nonlinearity is considered.
Such families can be found using the Newton method and they are displayed in Fig.~\ref{fig3}. Such families are characterized by the dependence of the power $P=\iint |\psi|^2 dx dy$ of the nonlinear state on its propagation constant $b$ that is now free parameter defining the family [in Fig.~\ref{fig3} red lines show the families of the disclination states, while blue lines show families of zero-energy states]. The case of pentagonal core is illustrated in Fig.~\ref{fig3}(a), while the case of heptagonal core in Fig.~\ref{fig3}(c). We found nonlinear states not only in the forbidden gaps, but also in the bulk band shown as gray region (i.e. nonlinear can drive propagation constant of nonlinear modes into bulk bands and continuation of the same family can be found in higher gap, as illustrated in the figure). One finds that the power $P$ of the nonlinear states increases with increase of the propagation constant $b$, changing nearly linearly within forbidden gaps. The shape of the topological disclination or zero-energy state strongly depends on the location of the propagation constant in the spectrum: in particular, the state gradually acquires long tails when the propagation constant penetrates into the band, especially when the slope of $P(b)$ dependence changes and the power starts increasing rapidly. Transformation of field modulus distributions in the nonlinear disclination (top row) and zero-energy (bottom row) states is illustrated in Figs.~\ref{fig3}(b) and \ref{fig3}(d) for pentagonal and heptagonal lattices, respectively. We choose four nonlinear states for each case, and two of them have propagation constants falling into forbidden gap. The corresponding in-gap states are well-localized. Only the in-band high-amplitude states corresponding to circles 4 couple with the bulk modes thereby gradually extending across the entire lattice. Notice that both disclination and zero-energy nonlinear families cross very narrow regions associated with the bands of linear edge and corner states [left outermost narrow gray region and vertical dashed lines in Figs.~\ref{fig3}(a) and \ref{fig3}(c)] practically without coupling with these states, 
which indicate that they belong to the class of embedded solitons~\cite{champneys.pd.152.340.2001}.
In the case of disclination states this is explained by the fact that they form at the disclination core located very far from the outer edges or corners of the structure, while in the case of zero-energy states very weak coupling with corner state can be explained by different symmetries of these states. Nevertheless, the weak coupling can result in some cases in the instability of the nonlinear zero-energy states, as will be shown later.

Among the most important properties of the nonlinear topological states is their stability. We found that stability properties can be very different in the lattices with pentagonal and heptagonal cores. We employ linear stability analysis method to investigate stability of the nonlinear disclination states. The method is based on substitution of perturbed solutions into governing Eq.~(\ref{eq1}), linearization around stationary state and solution of the resulting linear eigenvalue problem that yields perturbation growth rates $\beta$  (see \textbf{Methods}). Nonlinear states are stable only if real part of the growth rate $\beta_{\rm re}=\rm{Re}\{\beta\}$ is zero or negative for all possible perturbations. We calculate perturbation growth rates for all nonlinear disclination and zero-energy families from Figs.~\ref{fig3}(a) and \ref{fig3}(c). Maximal real part of the perturbation growth rate $\max\{\beta_{\rm re}\}$ versus $b$ is plotted in Figs.~\ref{fig4}(a,\,b) and \ref{fig4}(c,\,d) for the disclination and the zero-energy state, respectively. In accordance with this analysis, stable branches were plotted with solid lines in Figs.~\ref{fig3}(a) and \ref{fig3}(c), while unstable ones were plotted with dashed lines. One finds that the in-gap nonlinear disclination states in the pentagonal disclination lattice are unstable in the largest part of the gap, while those in the heptagonal disclination lattice are stable. Interestingly, when nonlinear disclination states couple with bulk modes in pentagonal disclination lattice, they become stable. In contrast, in heptagonal disclination lattice such states destabilize when they enter the bulk band. As shown in Figs.~\ref{fig4}(c) and \ref{fig4}(d) the nonlinear zero-energy states may be unstable in the vicinity of the edge/corner bands, but they become completely stable with increase of the propagation constant $b$. When they penetrate into the bulk band, nonlinear zero-energy states remain stable in the pentagonal case, but become unstable in the heptagonal case.

Last but not least, we investigate the possibility of excitation of the nonlinear disclination states by Gaussian input beams. To this end, we launch a Gaussian beam into one of the sites of the disclination core. By adjusting its power, we record the corresponding propagation dynamics. To monitor transformation of the beam during propagation, we calculate the form-factor of the field distribution at different distances that can be defined as:
\begin{equation}\label{eq7}
f^2 = \left. {\iint|\psi|^4 dx dy} \right/ {\left(\iint |\psi|^2 dx dy\right)^2} .
\end{equation}
The quantity $f^{-1}$ characterizes the width of the wavepacket. We first investigate the case of the pentagonal disclination lattice. Corresponding results illustrating excitation dynamics are shown in Figs.~\ref{fig5}(a) and \ref{fig5}(b). When the power of the Gaussian beam is low, $P=0.01$, the beam does not remain exclusively in the excited site, the power couples between all five disclination sites resulting in the oscillations of beam width upon propagation, see black curve in Fig.~\ref{fig5}(a) and field modulus distributions at different distances shown in the first row of Fig.~\ref{fig5}(b). In this case the input Gaussian excites superposition of several disclination states and they experience subsequent beatings, with excitation remaining always around disclination core. Increasing the power of the Gaussian beam to $P=0.09$ leads to strong localization in the only excited site, i.e. excitation of a strongly localized variant of the nonlinear disclination state, when nonlinearity suppresses coupling also between sites of the disclination core, see the red curve in Fig.~\ref{fig5}(a) and second row in Fig.~\ref{fig5}(b). Similar results can be obtained if we launch a Gaussian beam into the disclination core of the heptagonal disclination lattice, as shown in Figs.~\ref{fig5}(c) and \ref{fig5}(d). According to the results of Fig.~\ref{fig5}, dynamical excitation of the nonlinear disclination states can be achieved at very low power levels (for example, nonlinear disclination state in the heptagonal lattice is excited even if $P=0.0081$), i.e. they are thresholdless, which is advantageous for potential experimental implementation.

\begin{figure}[htpb]
\centering
\includegraphics[width=\columnwidth]{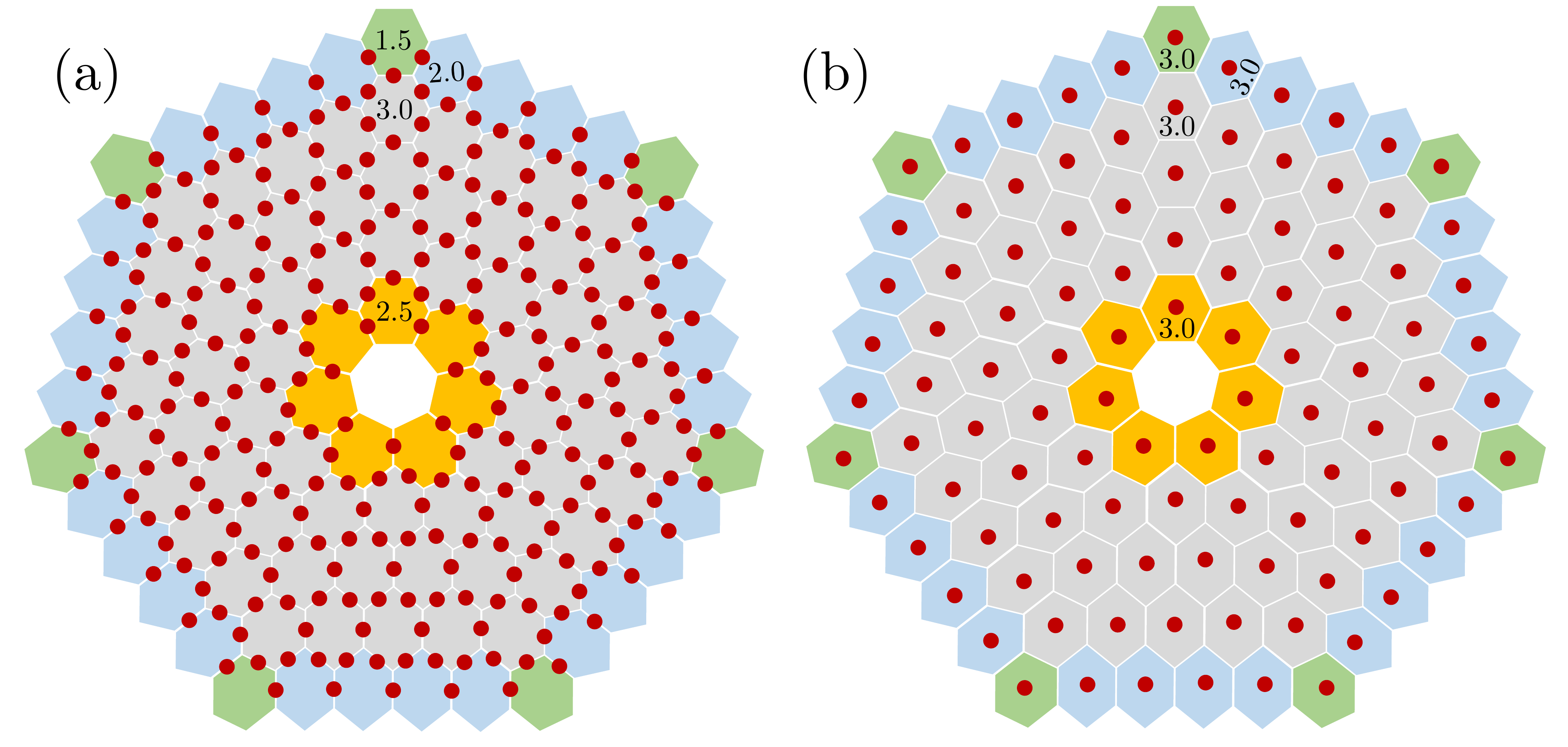}
\caption{\textbf{Wannier center (red dots) and spectral charge per unit cell in the disclination lattice with a heptagonal core.}
(a) Topological insulator phase with $r>r_c$. (b) Normal insulator phase with $r<r_c$.
The unit cell with same color has same spectral charge.}
\label{fig6}
\end{figure}

\section{Conclusion}

Summarizing, we have investigated nonlinear disclination states in disclination lattices with a pentagonal/heptagonal cores that are constructed by removing/inserting a sector with the Frank angle $2\pi/6$ from/into the honeycomb lattice. The disclination lattice enters into topological insulator phase upon variation of the separation among lattice sites in each unit cell, this is accompanied by the shift of the Wannier centers to the boundaries of the unit cell. In this phase, there appears fractional charge in the unit cell around the disclination core due to filling anomaly, which predicts the existence of disclination states according to the bulk-disclination correspondence. We obtained rich nonlinear disclination state families bifurcating from such linear topological states. We have confirmed that such states are practically thresholdless in contrast to nonlinear self-sustained states in periodic photonic lattices that always require certain minimal power for their formation. The interesting property of the system considered here is that disclination states coexist in it with zero-energy, corner and edge states. Despite their coexistence, the bifurcation of nonlinear disclination states into the gap occurs in such a way that they do not show any noticeable coupling with other states (for example, with zero-energy or corner states) present in the topological gap, even though their propagation constants may overlap, as one can see in Fig.~\ref{fig3}. Moreover, while the phenomenon of bifurcation of localized gap states from defects is known even in trivial periodic structures\cite{sukhorukov.pre.63.036601.2001}, we would like to stress here that in our case, in contrast to trivial inclusion of local defect, the deformation of the lattice after removal or addition of the sector is global and affects the entire structure, in a sense that the structure becomes aperiodic, but still it can possess the gap in the spectrum, where topological disclination states may form. We have also found that stability of disclination states crucially depends on the geometry of the lattice: nonlinear disclination states in the lattice with pentagonal core are unstable in the largest part of the gap, while nonlinear disclination states in the lattice with heptagonal core are instead stable in the entire gap. Our results not only introduce the first example of the nonlinear topological disclination states in photonic system, but they also open a pathway to enhancement of light-matter interactions. Additionally, we believe that the work will inspire investigations of disclination states in non-Hermitian systems, realization of topological lasing in such states, and design of compact on-chip photonic devices.

\section{Methods}

\subsection{Topological invariant}
The topological invariant that can be associated with the band gap of the disclination lattice~\cite{benalcazar.prb.99.245151.2019, li.prb.101.115115.2020, peterson.nature.589.376.2021, liu.nature.589.381.2021, wu.pr.9.668.2021} is defined by
\begin{equation}
\chi=(\chi_{\rm M},\chi_{\rm K}),
\end{equation}
where the high symmetry indicators are $\chi_{\rm M}=\#{\rm M}_1^{(2)}-\#\Gamma_1^{(2)}$ and $\chi_{\rm K}=\#{\rm K}_1^{(3)}-\#\Gamma_1^{(3)}$. Here $\#\Pi_q^{(n)}$ is the number of bands below the bandgap at a high-symmetry point $\Pi = \Gamma, \,\rm M,\, K$ [see the inset in Fig.~\ref{fig1}(a)] with the eigenvalue of the $C_n$ rotation matrix $e^{i2\pi(q-1)/n}~(q=1,\cdots,n)$. It can be shown for the disclination lattices with pentagonal and heptagonal cores that $\chi=(2,\,0)$ for $r>r_c$ and $\chi=(0,\,0)$ for $r<r_c$, that explains the appearance of topologically nontrivial states. In addition, the bulk-disclination correspondence establishes the link between the fractional disclination charge and the localized states bound to the disclination core~\cite{liu.nature.589.381.2021, wu.pr.9.668.2021}. The fractional disclination charge $\mathcal{Q}$ is determined by the topological index of the gap~\cite{benalcazar.prb.99.245151.2019, li.prb.101.115115.2020, peterson.nature.589.376.2021, liu.nature.589.381.2021, wu.pr.9.668.2021} and can be written as
\begin{equation}
\mathcal{Q} = \frac{\Omega}{2\pi} \left(\frac{3}{2}\chi_{\rm M}-\chi_{\rm K} \right) \, {\rm modulo}\, 1,
\end{equation}
which means that the fractional charge is $\mathcal Q=1/2$ for $r>r_c$ and $\mathcal Q=0$ for $r<r_c$. 
Here $\Omega=2\pi/6$ is the Frank angle.
The fractional disclination charge can be also obtained by counting the number of the Wannier centers.
For example, as shown in the insets in spectra presented in Figs.~\ref{fig1}(d) and \ref{fig1}(f), the Wannier centers are located at the edges of the unit cell if $r>r_c$ and at its center if $r<r_c$. 
In Fig.~\ref{fig6}, we show the Wannier center and the spectral charge in each unit cell in the disclination lattice with a heptagonal core.
The unit cell around the disclination core has five bulk Wannier centers if $r>r_c$, which give a $5/2$ spectral charge per unit cell, as shown in Fig.~\ref{fig6}(a). While if $r<r_c$, the spectral charge per unit cell around the disclination core is $3$ [see Fig.~\ref{fig6}(b)]. 

\subsection{Linear stability analysis method}
To perform linear stability analysis of the obtained nonlinear disclination states, we write perturbed solution in the form
\begin{equation}\label{eq5}
\psi = \left[ u(x, y)+v(x, y) e^{\beta z}+w^{*}(x, y) e^{\beta^{*} z} \right] e^{i b z},
\end{equation}
where $v,w\ll 1$ are small perturbations, and $\beta$ is the perturbation growth rate that can be complex. Inserting Eq.~(\ref{eq5}) into Eq.~(\ref{eq1}) and linearizing it around stationary solution $u$, we arrive at the linear eigenvalue problem:
\begin{equation}\label{eq6}
\begin{split}
i \beta v= \,&-\frac{1}{2}\left(\frac{\partial}{\partial x^{2}}+\frac{\partial}{\partial y^{2}}\right) v-({\mathcal R}-b) v-2|u|^{2} v-|u|^{2} w,\\
i \beta w = \,& +\frac{1}{2}\left(\frac{\partial}{\partial x^{2}}+\frac{\partial}{\partial y^{2}}\right) w + ({\mathcal R}-b) w + 2|u|^{2} w + |u|^{2} v.\\
\end{split}	
\end{equation}
Solving the problem (\ref{eq6}) using standard eigenvalue solver, we obtain the dependence of the perturbation growth rate $\beta$ (and associated perturbation profiles $v,w$) on the propagation constant $b$ for a given family of the nonlinear states. If $\beta_{\rm{re}} \le 0$ for all possible perturbations, corresponding nonlinear state $u$ is linearly stable and in the presence of small perturbations it will exhibit only small amplitude oscillations upon evolution, otherwise, if at least one perturbation mode has ${\beta_{\rm{re}} > 0}$, the nonlinear state is unstable and will decay in the course of propagation.\\

\noindent\textbf{Acknowledgements}\\
This work was supported by the National Natural Science Foundation of China (12074308, U1537210), the Russian Science Foundation (21-12-00096), and the Fundamental Research Funds for the Central Universities (xzy022022058).
\\

\noindent\textbf{Data availability}\\
The data that supports the results within this paper are available from the corresponding
authors upon reasonable request.
\\

\noindent\textbf{Code availability}\\
The codes used to obtain the results described in this paper are available from the corresponding authors upon reasonable request.
\\

\noindent\textbf{Competing interests}
The authors declare no competing interests.
\\

\noindent\textbf{Correspondence} and requests for materials should be addressed to Y.Z.
\\

%

\end{document}